\newif{\ifiopformat}
\iopformatfalse

\newif{\ifhavepgf}
\havepgffalse

\ifiopformat
\documentclass[12pt,a4paper]{iopart}
\else
\documentclass[aps,prb,reprint,showpacs,showkeys,floatfix,a4paper]{revtex4-1}
\fi

\usepackage[utf8]{inputenc}
\usepackage[english]{babel}
\usepackage{csquotes}

\ifiopformat
\usepackage{iopams}
\expandafter\let\csname equation*\endcsname\relax
\expandafter\let\csname endequation*\endcsname\relax
\usepackage[numbers,square,sort&compress]{natbib}
\newcommand{\refcite}[1]{\cite{#1}}
\newcommand{\refscite}[1]{\cite{#1}}
\else
\providecommand{\submitto}{}
\providecommand{\SUST}{}
\usepackage{natbib}
\newcommand{\refcite}[1]{Ref. \onlinecite{#1}}
\newcommand{\refscite}[1]{Refs. \onlinecite{#1}}

\fi
\usepackage{natmove} %

\usepackage{amssymb}
\usepackage{mathtools}
\usepackage{commath}
\usepackage[mode=math,range-units=single]{siunitx}
\newcommand{\sod}[2]{\dif #1 /\! \dif #2}

\usepackage{xcolor}
\usepackage[caption=false]{subfig}

\ifhavepgf
\usepackage{pgfplots}
\usepackage{tikz}
\usetikzlibrary{plotmarks}
\usetikzlibrary{matrix}
\usetikzlibrary{calc}
\usetikzlibrary{external}
\pgfplotsset{compat=1.5.1}
\pgfplotsset{plot coordinates/math parser=false}
\tikzsetexternalprefix{tikz/}

\tikzset{%
  /tikz/external/only named = true,
}

\tikzset{%
  every pin edge/.append style = {<-,semithick},
  every pin/.append style = {font=\normalsize},
  subplotleft/.style = {baseline,
    trim right={($(current axis.south east) + (0.75em,0)$)},
    trim left={($(current axis.south west) - (3em,0)$)},
    every node/.prefix style = {inner sep=0.3333em}},
  subplotright/.style = {baseline,
    trim right={($(current axis.south east) + (0.75em,0)$)},
    trim left={($(current axis.south west) - (2.5em,0)$)},
    every node/.prefix style = {inner sep=0.3333em}},
  plotcontainer/.style = {inner sep=0, anchor=base east},
}

\pgfkeys{%
  /pgf/number format/set thousands separator={\,},
}
\pgfplotsset{%
  xlabel near ticks,
  ylabel near ticks,
  every axis plot post/.append style = semithick,
  axis line style = thin,
  tick style = thin,
  every pin edge/.append style = {<-,semithick},
  subplotaxis/.style = {
    tick label style = {font=\footnotesize},
    label style = {font=\small}, %
    tick label style = {inner xsep=0.125em},
    label style = {inner xsep=0.125em},
  },
}

\else

\usepackage{graphicx}
\usepackage{tikzexternal}
\tikzsetexternalprefix{tikz/}

\fi

\newlength{\figureheight}
\newlength{\figurewidth}
\ifiopformat
\else
\fi
\usepackage{ifpdf}
\ifpdf
\usepackage[pdftex]{hyperref} %
\fi
\usepackage[noabbrev]{cleveref} %
\crefname{equation}{\unskip}{\unskip}
\Crefname{equation}{Equation}{Equations}
\usepackage{microtype} %

\usepackage{array}
\usepackage{booktabs}

\ifhavepgf
\usepackage{pgfplotstable}
\pgfplotstableset{%
  include outfiles,
  every table/.append style={outfile={tikz/#1.tex}}, %
  every head row/.style = {before row=\toprule, after row=\midrule},
  every last row/.style = {after row=\bottomrule},
  row style/.style 2 args={
      every row #1 column 0/.style={#2},
      every row #1 column 1/.style={#2},
      every row #1 column 2/.style={#2},
      every row #1 column 3/.style={#2},
      every row #1 column 4/.style={#2},
      every row #1 column 5/.style={#2},
      every row #1 column 6/.style={#2},
      every row #1 column 7/.style={#2},
      every row #1 column 8/.style={#2},
      every row #1 column 9/.style={#2},
      every row #1 column 10/.style={#2},
  }
}
\else

\fi

\newcommand{\mytitle}{Broadband illumination of superconducting pair breaking photon detectors}
\newcommand{\tu}[1]{\ensuremath{_{\mathrm{#1}}}}

\ifpdf
\hypersetup{%
  pdfinfo={%
    Title={\mytitle},
    Author={T Guruswamy, D J Goldie and S Withington},
  }
}
\fi

\begin{document}
\title[]{\mytitle}
\ifiopformat
\author{T Guruswamy, D J Goldie and S Withington}
\ead{tg307@mrao.cam.ac.uk}
\address{Quantum Sensors Group, Cavendish Laboratory, University of Cambridge,
J J Thomson Avenue, Cambridge, CB3~0HE, UK}
\else
\author{T Guruswamy}
\email{tg307@mrao.cam.ac.uk}
\author{D J Goldie}
\author{S Withington}
\affiliation{Quantum Sensors Group, Cavendish Laboratory, University of Cambridge,
J J Thomson Avenue, Cambridge, CB3~0HE, UK}
\fi
\date{\today}

\begin{abstract}
Understanding the detailed behaviour of superconducting pair breaking photon detectors such as Kinetic Inductance Detectors requires knowledge of the nonequilibrium quasiparticle energy distributions.
We have previously calculated the steady state distributions resulting from uniform absorption of monochromatic sub gap and above gap frequency radiation by thin films.
In this work, we use the same methods to calculate the effect of illumination by broadband sources, such as thermal radiation from astrophysical phenomena or from the readout system.
Absorption of photons at multiple above gap frequencies is shown to not change the structure of the quasiparticle energy distribution close to the superconducting gap.
Hence for typical absorbed powers, we find the effects of absorption of broadband pair breaking radiation can simply be considered as the sum of the effects of absorption of many monochromatic sources.
Distribution averaged quantities, like quasiparticle generation efficiency $\eta$, match exactly a weighted average over the bandwidth of the source of calculations assuming a monochromatic source.
For sub gap frequencies, however, distributing the absorbed power across multiple frequencies does change the low energy quasiparticle distribution.
For moderate and high absorbed powers, this results in a significantly larger $\eta$ -- a higher number of excess quasiparticles for a broadband source compared to a monochromatic source of equal total absorbed power.
Typically in KIDs the microwave power absorbed has a very narrow bandwidth, but in devices with broad resonance characteristics (low quality factors), this increase in $\eta$ may be measurable.
\end{abstract}
\submitto{\SUST}
\maketitle

\section{Introduction}

Kinetic Inductance Detectors~\cite{Day2003,Vardulakis2008} (KIDs) and related superconducting resonator based devices~\cite{Zmuidzinas2012} have become key technologies in photon sensing and quantum computation~\cite{Schoelkopf2008}.
Absorbed photons can break Cooper pairs creating excess quasiparticles in a superconducting thin film. The detailed quasiparticle energy distribution determines the superconductor's electrical~\cite{Mattis1958} and thermal~\cite{Bardeen1959} properties, so in recent work we have developed a framework to calculate the steady state quasiparticle and phonon energy distributions in illuminated thin films~\cite{Goldie2013,Guruswamy2014,Guruswamy2015b}.

Due to their low noise and frequency range, KIDs are being developed for measurements of signals such as the Cosmic Microwave Background (CMB)~\cite{Day2003,Calvo2010}, which peaks at \SI{160}{GHz}.
As well as being close to the gap frequency of typical low critical temperature superconductors,
the detected CMB signal, even if limited by atmospheric windows, is broadband -- perhaps 30\% bandwidth.
KIDs may also experience broadband above gap illumination due to background loading, and broadband sub gap illumination due to thermal noise in the readout line.
An implicit assumption of existing models is that the response to broadband absorbed power is not different from the response to illumination at a single frequency, but to our knowledge no previous work investigates or quantifies this claim.

Our calculations thus far have included one monochromatic sub gap source and one monochromatic above gap source.
These calculations have shown that absorbed sub gap microwave radiation can also increase the excess quasiparticle population despite not directly breaking pairs~\cite{Goldie2013}.
The effect of both separate and simultaneous absorption of above gap (pair breaking) radiation has also been quantified, including calculating the quasiparticle generation efficiency $\eta$, which we define as the fraction of absorbed power which creates excess low energy quasiparticles.
We have shown $\eta$ to be frequency dependent, particularly near the superconducting gap frequency $2\Delta/h$, where $\Delta$ is the energy gap.
Using these results, we have been able to include quasiparticle heating and nonequilibrium effects in device models~\cite{Thompson2013}.
Experimental measurements have shown excess quasiparticles due to microwave power~\cite{DeVisser2013a} and, for frequencies close to the gap frequency, a frequency dependent detector response to above gap radiation~\cite{DeVisser2015a} in agreement with calculations.

For a thermal radiation source, the quasiparticle temperature must approach an equilibrium determined by the effective thermal conductance associated with the electron-phonon interaction and phonon escape time.
Here we instead consider sources with finite bandwidth illuminating Al thin films and driving the quasiparticle energy distribution to a nonequilibrium steady state. A typical CMB signal frequency of $\SI{175}{GHz}$ is $4\Delta/h$, so we consider above gap sources with centre frequencies $\nu_s$ of $2\Delta/h$ to $10\Delta/h$ and bandwidths $\delta \nu_s$ up to $8\Delta/h$. We also include sub gap sources at typical readout frequencies $\nu_p$ for microwave KIDs of $\SI{4}{GHz}$ ($0.09\Delta/h$) to $\SI{8}{GHz}$, with bandwidths $\delta \nu_p$ as multiples of our numerical energy bin width of $0.005\Delta$ ($\SI{0.2}{GHz} \times h$).
However our work here is independent of material and device design, as we have shown that when the parameters to our model are scaled appropriately (energies by $\Delta$, temperatures by $T_c$), the results are material independent~\cite{Guruswamy2015b}.
We then compare the nonequilibrium distributions to results for a monochromatic source of the same total absorbed power, seeking any changes to distribution-averaged quantities such as $\eta$ introduced by the broadband nature of the source.

\section{Methods}

The steady state nonequilibrium quasiparticle and phonon distributions were calculated by solving the Chang and Scalapino nonlinear kinetic equations~\cite{Chang1977,Chang1978}, which describe the rates of change of the quasiparticle and phonon distributions due to electron-phonon interactions in bulk superconductors, in an energy conserving way using Newton-Raphson iteration.
This method assumes BCS theory applies, in particular that weak coupling holds, for the clean, thin ($\SI{40}{nm} < d < \SI{500}{nm}$) superconducting films typical of KIDs.
Based on the measured properties of our own sputtered films, we estimate that electron-phonon interactions are dominant over electron-electron interactions for the relevant quasiparticle energies ($3\Delta$ to $\sim 20\Delta$) and commonly used low-$T_c$ superconductors Al, Mo, Ta, Nb, and NbN~\cite{Guruswamy2014,Guruswamy2015b}.
Our calculation fixes the superconducting energy gap $\Delta = \Delta(T_b)$ as constant ($T_b$ is the substrate or bath temperature), which we find to be a valid approximation for the absorbed powers and low reduced temperatures ($T_b \ll T_c$, typically $T_b \sim 0.1\,T_c$, where $T_c$ is the superconductor critical temperature) considered~\cite{DeVisser2013a}.
Further details of the method are given in~\refscite{Goldie2013,Guruswamy2014,Guruswamy2015b}.

If the superconducting thin film uniformly absorbs a constant flux of photons at a frequency $\nu$, we include in the Chang and Scalapino equations a quasiparticle injection term $I_{qp}(E) = B K_{qp}(E,\nu)$~\cite{Eliashberg1972,Goldie2013}, where the prefactor $B$ normalises the total power absorbed. $K_{qp}(E,\nu)$ is the drive term describing relative rates of quasiparticle generation as a function of quasiparticle energy $E$ given absorption of photons of frequency $\nu$. It is given by $K_p(E,\nu)$ for $\nu \le 2\Delta$ and $K_s(E,\nu)$ for $\nu > 2\Delta$, where
\begin{align}
\begin{split}
K_{p}(E,\nu) &= 
2\Bigg[ \rho(E + h\nu) \left( 1 + \frac{\Delta^2}{E(E+h\nu)} \right) \\
&\quad {}\times \left( f(E+h\nu) - f(E) \right) \\
&\quad {}- \rho(E - h\nu) \left( 1 + \frac{\Delta^2}{E(E-h\nu)} \right) \\
&\quad {}\times \left( f(E) - f(E-h\nu) \right) \Bigg]
\end{split} \label{eq:k_p} \\
\begin{split}
K_{s}(E,\nu) &=
K_{p}(E,\nu) + 2\rho(h\nu - E) \left( 1 - \frac{\Delta^2}{E(h\nu - E)} \right) \\
&\quad {}\times \left( 1 - f(E) - f(h\nu - E) \right)  \; .
\end{split}
\label{eq:k_s}
\end{align}
Here $\rho(E)$ is the broadened BCS quasiparticle density of states~\cite{Zmuidzinas2012,Guruswamy2014}, and $f(E)$ is the (possibly nonequilibrium) quasiparticle energy distribution at energy $E$. The final term in \cref{eq:k_s} is only nonzero if $h\nu > 2\Delta$, differentiating the drive term representing absorption of signal (above gap frequency) photons $K_{s}(E,\nu)$ from the term \cref{eq:k_p} representing absorption of probe or readout (sub gap frequency) photons $K_{p}(E,\nu)$.

A broadband photon source with a fixed bandwidth was included by modifying the quasiparticle injection term to sum over the frequency range of interest. For a superconductor absorbing photons with frequencies $\nu_1 \ldots \nu_N$, the quasiparticle injection term $I_{qp}(E)$ was included as
\begin{equation}
\begin{split}
I_{qp}(E) = B \sum_{i=1}^N w_i K_{qp}(E,\nu_i) \; ,
\label{eq:Iqp}
\end{split}
\end{equation}
where $K_{qp}(E,\nu_i)$ is either $K_s$ or $K_p$ as in \cref{eq:k_p,eq:k_s}, and $w_i$ is a weighting factor adjusting the fraction of total power absorbed from photons of frequency $\nu_i$.
All frequencies included were required to align with specific energy bins in our discretisation scheme~\cite{Guruswamy2015b}.

The weighting factor $w_i$ can in principle be any function of frequency, to force the power absorbed at each frequency $P(\nu_i)$ given by
\begin{equation}
P(\nu_i) = 4 \, N\!(0) \, B \, w_i \int_{E=\Delta}^{\infty} E \rho(E) K_{qp}(E, \nu_i) \dif E \; ,
\end{equation}
to have the desired frequency dependence. Only the relative values of $w_i$ are important, as the prefactor $B$ scales the expression to ensure the calculated absorbed power equals the specified absorbed power. Here we specified the absorbed power to be divided equally among the frequencies present, by setting the weighting factor to
\begin{equation}
  w_i = \left( \int_{E=\Delta}^{\infty} E \rho(E) K_{qp}(E,\nu_i) \dif E \right)^{-1} \; ,
\end{equation}
ensuring $P(\nu_i)$ had no frequency dependence.
The numerical method solves for the values of $B$, quasiparticle energy distribution $f(E)$, and phonon energy distribution which conserve energy such that $\sum_i P(\nu_i)$, quasiparticle-phonon energy flow, and phonon escape to substrate all equal the total specified absorbed power $P$.

\begin{figure}[htb]
  \centering
  \tikzsetnextfilename{letter_figure1}
  \input{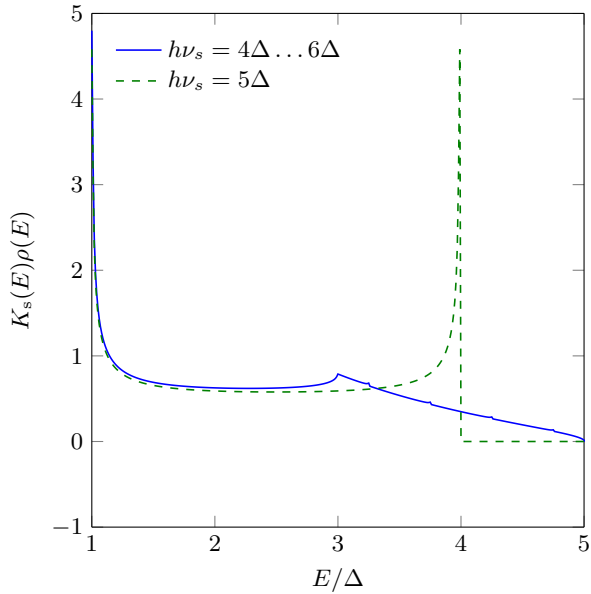}
  \caption{Sum of signal (pair breaking, above gap frequency) photon drive terms describing relative rate of generation of quasiparticles $K_{s}$ as a function of quasiparticle energy $E$,
    scaled by the quasiparticle density of states $\rho(E)$, in the broadband frequency (solid blue) and monochromatic (dashed green) case.
    Both assume absorption by a thermal quasiparticle distribution in Al at $T = 0.1\,T_c$.
    Broadband signal includes frequencies spaced by $0.01\Delta$ from $4\Delta$ to $6\Delta$; monochromatic signal is at $h\nu_s = 5\Delta$.}
  \label{fig:1}
\end{figure}

The quasiparticle injection term $I_{qp}$ represents the contribution to $\sod{f(E)}{t}$ due to quasiparticle-photon interactions; for the rate of change of number of quasiparticles at a given energy we need $\sod{N_{qp}(E)}{t} \propto \rho(E)\sod{f(E)}{t}$.
\Cref{fig:1} compares this contribution to quasiparticle number change from the broadband signal drive term ($\rho(E)\sum_i K_{s}(E,\nu_i)$, $h\nu_1 = 4\Delta$, $h\nu_{N} = 6\Delta$, bin width = $0.01\Delta$, solid blue) to $\rho(E) K_{s}(E,5\Delta/h)$ (dashed green), assuming a thermal quasiparticle distribution at $T = 0.1\,T_c$.
Terms $K_{s}(E,\nu)$ representing absorption of above gap photons by a thermal quasiparticle distribution have a low energy peak at energy $E = \Delta$ whose position is independent of frequency, and a higher energy peak at $E = h\nu - \Delta$ whose position is dependent on frequency. Therefore, the low energy structure of $K_{s}$ in the broadband case is the same as in the single frequency case, while the high energy structure is spread out over the range of frequencies absorbed into a step-like feature.

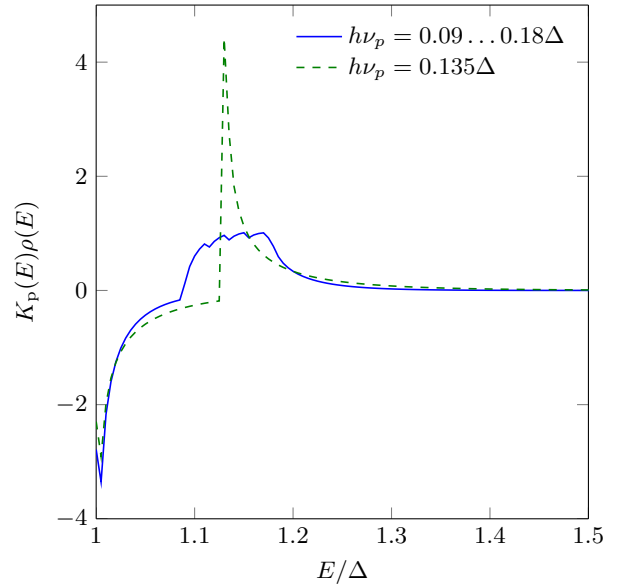
\begin{figure}[htb]
  \centering
  \tikzsetnextfilename{letter_figure11P}
%
%
\begin{tikzpicture}

\begin{axis}[%
width=0.95092\figurewidth,
height=\figureheight,
at={(0\figurewidth,0\figureheight)},
scale only axis,
separate axis lines,
every outer x axis line/.append style={black},
every x tick label/.append style={font=\color{black}},
xmin=1,
xmax=1.5,
xlabel={$E / \Delta$},
every outer y axis line/.append style={black},
every y tick label/.append style={font=\color{black}},
ymin=-4,
ymax=5,
ylabel={$K\tu{p}(E) \rho(E)$},
legend style={legend cell align=left,align=left,fill=none,draw=none}
]
\addplot [color=blue,solid]
  table[row sep=crcr]{%
1	-2.78594262724856\\
1.005	-3.35476403205505\\
1.01	-2.16867953433494\\
1.015	-1.60080815938329\\
1.02	-1.25188062044329\\
1.025	-1.01120615028664\\
1.03	-0.833948224340471\\
1.035	-0.69782188223288\\
1.04	-0.590229802084246\\
1.045	-0.503392680946813\\
1.05	-0.432187634348509\\
1.055	-0.373073123440465\\
1.06	-0.323506428428928\\
1.065	-0.28160611346851\\
1.07	-0.245945602966115\\
1.075	-0.215421214274591\\
1.08	-0.189164603257017\\
1.085	-0.16648282173365\\
1.09	0.119848670995057\\
1.095	0.420248444762932\\
1.1	0.602886568920802\\
1.105	0.725246148659582\\
1.11	0.814099696710486\\
1.115	0.761818828956184\\
1.12	0.86007467308669\\
1.125	0.924257755868504\\
1.13	0.965942689271455\\
1.135	0.88611449304813\\
1.14	0.951024290284899\\
1.145	0.989590503541536\\
1.15	1.01121500600064\\
1.155	0.923331239348407\\
1.16	0.971676086238869\\
1.165	0.997632361486545\\
1.17	1.00946468759405\\
1.175	0.92068091211078\\
1.18	0.774184856165443\\
1.185	0.592886411685487\\
1.19	0.477707045664663\\
1.195	0.397358638814948\\
1.2	0.336030067160113\\
1.205	0.287205325014687\\
1.21	0.247350252122995\\
1.215	0.214265931442332\\
1.22	0.186466297701337\\
1.225	0.162890051129564\\
1.23	0.14274884040345\\
1.235	0.125439632240527\\
1.24	0.110490588050873\\
1.245	0.0975258388848791\\
1.25	0.0862415843667959\\
1.255	0.076389315996797\\
1.26	0.0677637061296651\\
1.265	0.0601936567868574\\
1.27	0.0535355503069902\\
1.275	0.0476680722464643\\
1.28	0.0424881809831432\\
1.285	0.0379079292533158\\
1.29	0.0338519289897163\\
1.295	0.0302553089481855\\
1.3	0.0270620546767411\\
1.305	0.024223648537451\\
1.31	0.0216979476240985\\
1.315	0.0194482520409915\\
1.32	0.0174425267805991\\
1.325	0.0156527484772094\\
1.33	0.0140543543852998\\
1.335	0.0126257755666593\\
1.34	0.0113480398447463\\
1.345	0.010204432866636\\
1.35	0.00918020779644523\\
1.355	0.00826233589158057\\
1.36	0.00743929158981417\\
1.365	0.00670086683976969\\
1.37	0.00603801029928945\\
1.375	0.00544268775056178\\
1.38	0.00490776067251057\\
1.385	0.00442688039661676\\
1.39	0.0039943956729541\\
1.395	0.00360527180516735\\
1.4	0.00325501978933881\\
1.405	0.00293963412244765\\
1.41	0.00265553813963303\\
1.415	0.00239953590230937\\
1.42	0.00216876979666982\\
1.425	0.00196068311855987\\
1.43	0.00177298701962115\\
1.435	0.00160363127386858\\
1.44	0.00145077839583382\\
1.445	0.00131278070303158\\
1.45	0.00118815996839598\\
1.455	0.00107558935383151\\
1.46	0.000973877355244356\\
1.465	0.000881953523299256\\
1.47	0.000798855753469505\\
1.475	0.000723718964371504\\
1.48	0.000655765005458654\\
1.485	0.000594293654362302\\
1.49	0.00053867458091015\\
1.495	0.000488340169464453\\
1.5	0.000442779103993238\\
1.505	0.000401530631465135\\
};
\addlegendentry{$h\nu_p = 0.09 \ldots 0.18 \Delta$};

\addplot [color=black!50!green,dashed]
  table[row sep=crcr]{%
1	-2.29641052197916\\
1.005	-2.90888564839515\\
1.01	-1.97791824698309\\
1.015	-1.53556248588909\\
1.02	-1.26292696690641\\
1.025	-1.07279823636089\\
1.03	-0.930378652781233\\
1.035	-0.818633689088904\\
1.04	-0.728073099118952\\
1.045	-0.652914037588165\\
1.05	-0.589392814488972\\
1.055	-0.53493305428314\\
1.06	-0.487699352321813\\
1.065	-0.44634125013345\\
1.07	-0.409838266875058\\
1.075	-0.377401851182201\\
1.08	-0.348410999059877\\
1.085	-0.322368620666165\\
1.09	-0.298871152905205\\
1.095	-0.277586889260072\\
1.1	-0.258240201152438\\
1.105	-0.2405998355027\\
1.11	-0.224470091830792\\
1.115	-0.209684071754796\\
1.12	-0.196098445161571\\
1.125	-0.183589343267524\\
1.13	4.42077194341007\\
1.135	2.74750200421964\\
1.14	1.82640786391398\\
1.145	1.39320599934991\\
1.15	1.12906949721266\\
1.155	0.946842213783307\\
1.16	0.811777618199055\\
1.165	0.706886912568607\\
1.17	0.62272088819529\\
1.175	0.553533630261529\\
1.18	0.49559477043479\\
1.185	0.446358072573775\\
1.19	0.404015455900952\\
1.195	0.367241286972927\\
1.2	0.335037685655296\\
1.205	0.306636715833005\\
1.21	0.281436211859987\\
1.215	0.258956328048008\\
1.22	0.238809311819086\\
1.225	0.22067797505685\\
1.23	0.204300040554794\\
1.235	0.189456550440016\\
1.24	0.175963141781835\\
1.245	0.163663383832937\\
1.25	0.152423622535431\\
1.255	0.142128943673739\\
1.26	0.132679977653409\\
1.265	0.123990345407053\\
1.27	0.115984598280924\\
1.275	0.108596542524921\\
1.28	0.101767866123688\\
1.285	0.0954470054254432\\
1.29	0.0895882035377574\\
1.295	0.0841507232593719\\
1.3	0.0790981854372631\\
1.305	0.0743980098023492\\
1.31	0.0700209400588147\\
1.315	0.0659406386491017\\
1.32	0.0621333394560015\\
1.325	0.0585775489303048\\
1.33	0.0552537878913833\\
1.335	0.0521443676465009\\
1.34	0.0492331951933473\\
1.345	0.0465056031705196\\
1.35	0.0439482009491519\\
1.355	0.0415487438514774\\
1.36	0.0392960179666206\\
1.365	0.0371797384319219\\
1.37	0.0351904593765564\\
1.375	0.0333194939964408\\
1.38	0.0315588434559857\\
1.385	0.0299011335015635\\
1.39	0.028339557830332\\
1.395	0.0268678273917025\\
1.4	0.0254801249116342\\
1.405	0.0241710640255981\\
1.41	0.0229356524873992\\
1.415	0.0217692589903994\\
1.42	0.0206675831970146\\
1.425	0.0196266286232401\\
1.43	0.0186426780687214\\
1.435	0.0177122713206221\\
1.44	0.016832184892167\\
1.445	0.0159994135850038\\
1.45	0.0152111536890872\\
1.455	0.0144647876551654\\
1.46	0.0137578700936022\\
1.465	0.01308811496958\\
1.47	0.012453383879014\\
1.475	0.0118516753020573\\
1.48	0.0112811147421076\\
1.485	0.0107399456679498\\
1.49	0.0102265211852545\\
1.495	0.0097392963712405\\
1.5	0.00927682121303552\\
1.505	0.00883773409623647\\
};
\addlegendentry{$h\nu_p = 0.135 \Delta$};

\end{axis}
\end{tikzpicture}%
  \caption{Sum of readout (sub gap frequency) photon drive terms describing relative rate of generation of quasiparticles $K_{p}$ as a function of quasiparticle energy $E$,
    scaled by the quasiparticle density of states $\rho(E)$, in the broadband frequency (solid blue) and monochromatic (dashed green) case.
    Both assume absorption by a thermal quasiparticle distribution in Al at $T = 0.1\,T_c$.
    Broadband readout includes frequencies spaced by $0.005\Delta$ from $0.09\Delta$ to $0.18\Delta$; monochromatic readout is at $0.135\Delta$.}
  \label{fig:11P}
\end{figure}

\Cref{fig:11P} compares the similar contribution to quasiparticle number change from a broadband readout drive term ($\sum_{i} K_{p}(E,\nu_i)$, $h\nu_1 = 0.09\Delta$, $h\nu_{N} = 0.18\Delta$, bin width $0.005\Delta$, solid blue) to $K_{p}(E, 0.135\Delta/h)$ (dashed green), again assuming a thermal quasiparticle distribution at $T = 0.1\,T_c$. Changing from the monochromatic to a broadband source, the sharp peak at $E = \Delta + h\nu_p$ is now spread out into a broad peak over the bandwidth of the source.
Significantly the changes in $K_{qp}(E)$ are at a much lower energy ($E \sim \Delta + h\nu_p$) than in the above gap case ($E \sim h\nu_s - \Delta$).

Once the quasiparticle energy distribution is calculated, we identify the quasiparticle generation efficiency $\eta$ as the appropriate measure of the change in overall quasiparticle number due to the absorbed power. By solving a modified set of Rothwarf-Taylor equations~\cite{Rothwarf1967} we find
\begin{equation}
  \eta = \left< E_{qp} \right> \frac{(N^2 - N_t^2)}{P_{abs}} \frac{2R}{1+\beta \tau_l} \; ,
\end{equation}
where $\left< E_{qp} \right>$ is the average energy of the excess quasiparticles, usually $\sim \Delta$; $N$ is the total quasiparticle number for the driven distribution; $N_t$ is the number of quasiparticles in the equilibrium distribution, $P_{abs}$ is the absorbed power; $R$ and $\beta$ are the distribution averaged quasiparticle recombination and phonon pair breaking rates respectively~\cite{Kaplan1976,Chang1977}; and $\tau_l$ is the phonon escape time from the thin film into the substrate, given as a ratio to the characteristic phonon lifetime~\cite{Kaplan1976}, $\tau_0^\phi$.
Along with the appropriate material dependent constant $\Sigma_s$, once known, $\eta$ can be used to calculate changes in total quasiparticle number (or effective temperature) due to absorbed power, and thereby include nonequilibrium heating effects in higher level device models without full calculation of the quasiparticle distributions~\cite{Goldie2013,Guruswamy2015b}. By itself, it can be considered as the proportion of absorbed power which supports the population of detectable excess low-energy quasiparticles. The remaining absorbed power is lost to the substrate as escaped phonons without contributing to increased quasiparticle number.
For a detector, after any electromagnetic absorption efficiencies are considered, a larger $\eta$ corresponds to a greater efficiency in measuring absorbed power.

\section{Results}
\subsection{Pair breaking (above gap frequency) photons}

\begin{figure}[htb]
  \centering
  \tikzsetnextfilename{letter_figure2}
  \input{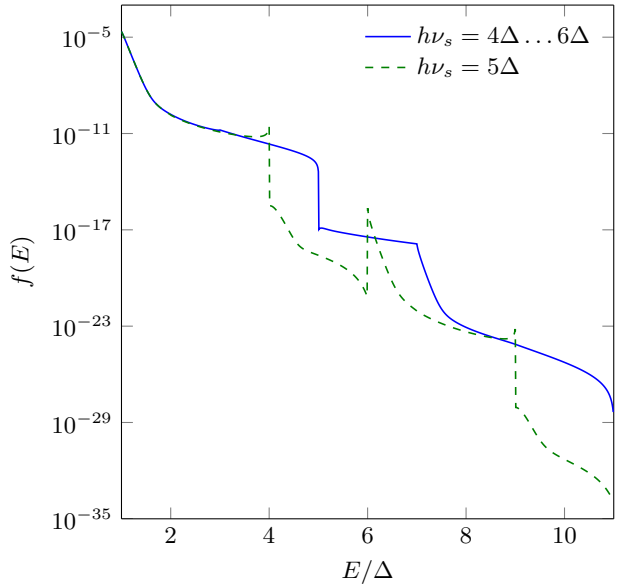}
  \caption{Steady state quasiparticle distribution $f(E)$ under illumination by monochromatic (dashed green) or broadband (solid blue) signal photons.
  Calculated for Al, with $P_{p} = 0$, $P_{s} = \SI{2}{W.m^{-3}}$, $\tau_l/\tau_0^\phi = 1$ and $T_b/T_c = 0.1$.}
  \label{fig:2}
\end{figure}

Using the quasiparticle injection term \cref{eq:Iqp}, we are able to calculate steady state quasiparticle and phonon distributions for thin films uniformly absorbing photons over a range of frequencies. \Cref{fig:2} shows the calculated steady state quasiparticle energy distribution $f(E)$ in the cases of absorbed pair breaking (above gap frequency) photons. For a monochromatic source (dashed green), the distribution has peaks at $h\nu - \Delta$ and $h\nu + \Delta$. For a broadband source (solid blue), these are instead extended over the bandwidth of the source into a step, however $f(E)$ near the superconducting energy gap $\Delta$ is unchanged.

Distribution-averaged quantities are largely determined by the low energy $f(E)$, close to the superconducting gap, due to the quasiparticle density of states peak at $E = \Delta$ and the exponential nature of the Fermi distribution. As a broadband above gap source does not introduce any new structure to the low energy $f(E)$ in comparison to a monochromatic source, quantities such as quasiparticle number and lifetime for thin films absorbing broadband pair breaking signals can be calculated by averaging (over the bandwidth of the source) the same quantities calculated for a monochromatic source of equal power.

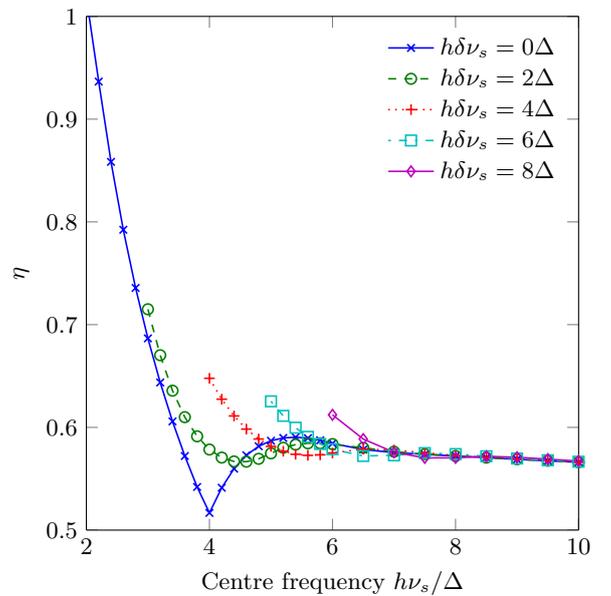
\begin{figure}[htb]
  \centering
  \tikzsetnextfilename{letter_figure3}
%
%
\definecolor{mycolor1}{rgb}{0.00000,0.75000,0.75000}%
\definecolor{mycolor2}{rgb}{0.75000,0.00000,0.75000}%
\begin{tikzpicture}

\begin{axis}[%
width=0.95092\figurewidth,
height=\figureheight,
at={(0\figurewidth,0\figureheight)},
scale only axis,
separate axis lines,
every outer x axis line/.append style={black},
every x tick label/.append style={font=\color{black}},
xmin=2,
xmax=10,
xlabel={Centre frequency $h\nu_s / \Delta$},
every outer y axis line/.append style={black},
every y tick label/.append style={font=\color{black}},
ymin=0.5,
ymax=1,
ylabel={$\eta$},
legend style={legend cell align=left,align=left,fill=none,draw=none}
]
\addplot [color=blue,solid,mark=x,mark options={solid}]
  table[row sep=crcr]{%
2	1.01855891317766\\
2.2	0.936572905453768\\
2.4	0.858413456015149\\
2.6	0.792334343906214\\
2.8	0.735709288005579\\
3	0.68663888835348\\
3.2	0.643704730580091\\
3.4	0.605823316949181\\
3.6	0.57215232846576\\
3.8	0.542027614210552\\
4	0.516749437543816\\
4.2	0.541276760723912\\
4.4	0.559810100677226\\
4.6	0.572887947827943\\
4.8	0.581628248473494\\
5	0.586982679042382\\
5.2	0.589710147525912\\
5.4	0.590403299824684\\
5.6	0.589521759923507\\
5.8	0.58742187566807\\
6	0.584382700202637\\
6.5	0.578429686500362\\
7	0.575510087889242\\
7.5	0.573723445440607\\
8	0.572171170937058\\
8.5	0.570534060845316\\
9	0.568925201483544\\
9.5	0.567425904112849\\
10	0.566049585654427\\
};
\addlegendentry{$h\delta\nu_s = 0 \Delta$};

\addplot [color=black!50!green,dashed,mark=o,mark options={solid}]
  table[row sep=crcr]{%
3	0.714981335426851\\
3.2	0.670234928043298\\
3.4	0.635741639068919\\
3.6	0.60994202866283\\
3.8	0.591298023097551\\
4	0.578442492190957\\
4.2	0.570619888960549\\
4.4	0.566834762382865\\
4.6	0.566626040215411\\
4.8	0.569414238337978\\
5	0.574720925361692\\
5.2	0.580185767077801\\
5.4	0.583229001566687\\
5.6	0.58449712884388\\
5.8	0.584527674217202\\
6	0.58373508536759\\
6.5	0.580084604141707\\
7	0.576380339400568\\
7.5	0.573931789172566\\
8	0.572148325900942\\
8.5	0.570546610809028\\
9	0.568993563239021\\
9.5	0.567507479835466\\
10	0.566124385997081\\
};
\addlegendentry{$h\delta\nu_s = 2 \Delta$};

\addplot [color=red,dotted,mark=+,mark options={solid}]
  table[row sep=crcr]{%
4	0.647768932814062\\
4.2	0.627435477317769\\
4.4	0.6111308996943\\
4.6	0.598340882391927\\
4.8	0.588665029636865\\
5	0.581516873145463\\
5.2	0.57672783954608\\
5.4	0.573845871614479\\
5.6	0.572764263290589\\
5.8	0.573266377549996\\
6	0.575105912587582\\
6.5	0.578866391232241\\
7	0.577950400496018\\
7.5	0.575356156457982\\
8	0.572710308261732\\
8.5	0.570723710778926\\
9	0.569136971426581\\
9.5	0.567699438486559\\
10	0.566334846860122\\
};
\addlegendentry{$h\delta\nu_s = 4 \Delta$};

\addplot [color=mycolor1,dash pattern=on 1pt off 3pt on 3pt off 3pt,mark=square,mark options={solid}]
  table[row sep=crcr]{%
5	0.625346633514137\\
5.2	0.611219383016122\\
5.4	0.599695059781332\\
5.6	0.590781481257739\\
5.8	0.583854389722934\\
6	0.578694171553327\\
6.5	0.572188991834258\\
7	0.572796870919395\\
7.5	0.5750004081342\\
8	0.574014523467651\\
8.5	0.571883901252306\\
9	0.569718116992273\\
9.5	0.568016203789309\\
10	0.566616324844353\\
};
\addlegendentry{$h\delta\nu_s = 6 \Delta$};

\addplot [color=mycolor2,solid,mark=diamond,mark options={solid}]
  table[row sep=crcr]{%
6	0.612244194795784\\
6.5	0.588633964586809\\
7	0.575759031204622\\
7.5	0.570329150752982\\
8	0.570326351538686\\
8.5	0.571841780985484\\
9	0.570909030548759\\
9.5	0.5690915108382\\
10	0.56724009840686\\
};
\addlegendentry{$h\delta\nu_s = 8 \Delta$};

\end{axis}
\end{tikzpicture}%
  \caption{Quasiparticle generation efficiency $\eta$ as a function of centre signal frequency $h\nu_s$, with varying bandwidth of signal $h\delta \nu_s$.
  Calculated for Al, with $P_{p} = 0$, $P_{s} = \SI{2}{W.m^{-3}}$, $\tau_l/\tau_0^\phi = 10$ and $T_b/T_c = 0.1$.}
  \label{fig:3}
\end{figure}

\Cref{fig:3} shows the calculated quasiparticle generation efficiency $\eta$. In the single frequency case (blue $\times$), the frequency dependence of $\eta$ shows a maximum of unity at $h\nu_s = 2\Delta$, a local minimum at $h\nu_s = 4\Delta$, followed by an increase dependent on the phonon trapping factor~\cite{Guruswamy2014}. When instead including a broadband source, the broadband $\eta$ matches an average of the monochromatic $\eta$ over the bandwidth of the source, smoothing out the minimum at $h\nu_s = 4\Delta$.

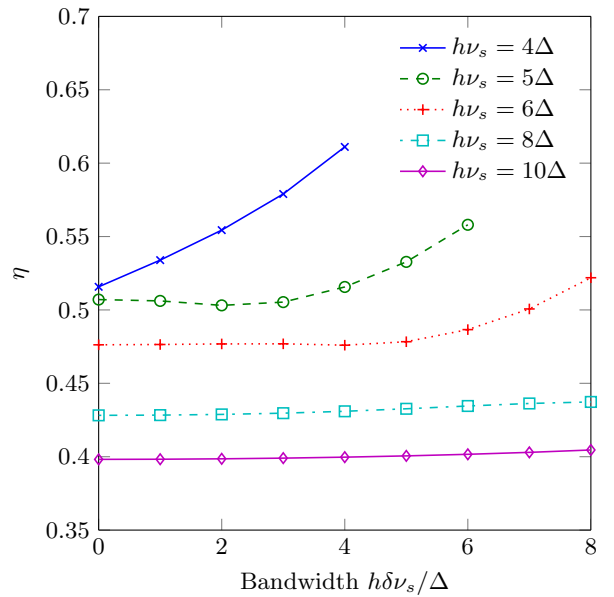
\begin{figure}[htb]
  \centering
  \tikzsetnextfilename{letter_figure4}
%
%
\definecolor{mycolor1}{rgb}{0.00000,0.75000,0.75000}%
\definecolor{mycolor2}{rgb}{0.75000,0.00000,0.75000}%
\begin{tikzpicture}

\begin{axis}[%
width=0.95092\figurewidth,
height=\figureheight,
at={(0\figurewidth,0\figureheight)},
scale only axis,
separate axis lines,
every outer x axis line/.append style={black},
every x tick label/.append style={font=\color{black}},
xmin=0,
xmax=8,
xlabel={Bandwidth $h\delta\nu_s / \Delta$},
every outer y axis line/.append style={black},
every y tick label/.append style={font=\color{black}},
ymin=0.35,
ymax=0.7,
ylabel={$\eta$},
legend style={legend cell align=left,align=left,fill=none,draw=none}
]
\addplot [color=blue,solid,mark=x,mark options={solid}]
  table[row sep=crcr]{%
0	0.515753166783631\\
1	0.533926984003759\\
2	0.55437370464095\\
3	0.578975818583363\\
4	0.611062498203597\\
};
\addlegendentry{$h\nu_s = 4 \Delta$};

\addplot [color=black!50!green,dashed,mark=o,mark options={solid}]
  table[row sep=crcr]{%
0	0.507055621528693\\
1	0.506139086801912\\
2	0.503108081696914\\
3	0.505320796313841\\
4	0.51564439489924\\
5	0.532733071066138\\
6	0.558009995038614\\
};
\addlegendentry{$h\nu_s = 5 \Delta$};

\addplot [color=red,dotted,mark=+,mark options={solid}]
  table[row sep=crcr]{%
0	0.476231120126487\\
1	0.476447805655037\\
2	0.476831770709508\\
3	0.476908578037251\\
4	0.475980852186089\\
5	0.47833811494606\\
6	0.486590485016391\\
7	0.50066781174825\\
8	0.521933362654968\\
};
\addlegendentry{$h\nu_s = 6 \Delta$};

\addplot [color=mycolor1,dash pattern=on 1pt off 3pt on 3pt off 3pt,mark=square,mark options={solid}]
  table[row sep=crcr]{%
0	0.428179348200676\\
1	0.428333735598961\\
2	0.428812936025128\\
3	0.429658050804609\\
4	0.430936517030391\\
5	0.432652565829287\\
6	0.434548932464575\\
7	0.436267050133757\\
8	0.437314626911962\\
};
\addlegendentry{$h\nu_s = 8 \Delta$};

\addplot [color=mycolor2,solid,mark=diamond,mark options={solid}]
  table[row sep=crcr]{%
0	0.398184334587154\\
1	0.398277752550543\\
2	0.398557782007574\\
3	0.399027801059054\\
4	0.399692103438217\\
5	0.400555592590663\\
6	0.401635927351303\\
7	0.402957721227332\\
8	0.404581586826734\\
};
\addlegendentry{$h\nu_s = 10 \Delta$};

\end{axis}
\end{tikzpicture}%
  \caption{Quasiparticle generation efficiency $\eta$ as a function of signal bandwidth $h\delta \nu_s$, with varying centre frequency $h\nu_s$.
  Calculated for Al, with $P_{p} = 0$, $P_{s} = \SI{2}{W.m^{-3}}$, $\tau_l/\tau_0^\phi = 1$ and $T_b/T_c = 0.1$.}
  \label{fig:4}
\end{figure}

\Cref{fig:4} shows that $\eta$ varies with signal bandwidth only if that bandwidth covers the significant features in the frequency dependence of $\eta$, i.e. $h\nu_s = 2\Delta \ldots 6\Delta$. Little dependence on bandwidth is shown at high frequencies ($h\nu_s \ge 10\Delta$), where $\eta$ is relatively constant with frequency.

\begin{figure}[htb]
  \centering
  \tikzsetnextfilename{letter_figure5}
%
%
\begin{tikzpicture}

\begin{axis}[%
width=0.95092\figurewidth,
height=\figureheight,
at={(0\figurewidth,0\figureheight)},
scale only axis,
separate axis lines,
every outer x axis line/.append style={black},
every x tick label/.append style={font=\color{black}},
xmin=2.5,
xmax=10,
xlabel={Centre frequency $h\nu_s / \Delta$},
every outer y axis line/.append style={black},
every y tick label/.append style={font=\color{black}},
ymin=-3.7,
ymax=-3,
ylabel={$10^3 \delta \sigma_2 / \sigma_n$ at \SI{4}{GHz}},
legend style={at={(0.97,0.03)},anchor=south east,legend cell align=left,align=left,fill=none,draw=none}
]
\addplot [color=blue,solid,mark=x,mark options={solid}]
  table[row sep=crcr]{%
2.63199492943499	-3.77\\
2.8	-3.65385573592647\\
3	-3.52961070256441\\
3.2	-3.41719695364162\\
3.4	-3.31485971985046\\
3.6	-3.22118081847123\\
3.8	-3.13500935765632\\
4	-3.0604639491898\\
4.2	-3.13346622890975\\
4.4	-3.18809776845796\\
4.6	-3.22670105743583\\
4.8	-3.25282790816317\\
5	-3.26935240717319\\
5.2	-3.2784843521938\\
5.4	-3.2819044757133\\
5.6	-3.28089234324125\\
5.8	-3.27642508253945\\
6	-3.26926257001503\\
6.5	-3.25560408484193\\
7	-3.25016498540975\\
7.5	-3.24769946102776\\
8	-3.24570108048761\\
8.5	-3.24327734588081\\
9	-3.24075718963002\\
9.5	-3.23838635653573\\
10	-3.23621893938508\\
};
\addlegendentry{$h\delta\nu_s = 0$ (monochromatic)};

\addplot [color=black!50!green,dashed]
  table[row sep=crcr]{%
3	-3.58661175022945\\
3.2	-3.48412268512988\\
3.4	-3.39936209313407\\
3.6	-3.3338691669158\\
3.8	-3.28484274547947\\
4	-3.24988789741099\\
4.2	-3.22705822919289\\
4.4	-3.20474883233857\\
4.6	-3.21167096969077\\
4.8	-3.21669317551709\\
5	-3.22889801299908\\
5.2	-3.24574141938427\\
5.4	-3.25821644054969\\
5.6	-3.26571714237645\\
5.8	-3.27059415298692\\
6	-3.27026128763919\\
6.5	-3.26334142117446\\
7	-3.25368643635926\\
7.5	-3.24848939152389\\
8	-3.24552001249145\\
8.5	-3.24316428670812\\
9	-3.24086818238101\\
};
\addlegendentry{moving average over $2\Delta$};

\addplot [color=red,dotted,mark=o,mark options={solid}]
  table[row sep=crcr]{%
3	-3.60133276222996\\
3.2	-3.4868810912414\\
3.4	-3.39600925613581\\
3.6	-3.32652307812964\\
3.8	-3.27561194419701\\
4	-3.24032996922341\\
4.2	-3.21906720871823\\
4.4	-3.20927023302175\\
4.6	-3.20973399450963\\
4.8	-3.21881435148441\\
5	-3.23506860250689\\
5.2	-3.25182372485955\\
5.4	-3.26174142595193\\
5.6	-3.26666901630546\\
5.8	-3.26810631466401\\
6	-3.26720328941832\\
6.5	-3.26006696460723\\
7	-3.25247650145144\\
7.5	-3.24816248072324\\
8	-3.24551672055406\\
8.5	-3.24319851269905\\
9	-3.24084556037008\\
9.5	-3.23852210850362\\
10	-3.23634493329195\\
};
\addlegendentry{$h\delta\nu_s = 2\Delta$};

\end{axis}
\end{tikzpicture}%
  \caption{Change in the imaginary part of the superconductor complex conductivity $\sigma_2$ at \SI{4}{GHz} as a function of centre signal frequency $h\nu_s$.
  An average (dashed green) of the monochromatic calculation (blue $\times$) over a bandwidth of $2\Delta$ matches the full calculation including a broadband signal with bandwidth $2\Delta$ (red $\bigcirc$).
  Calculated for Al, with $P_{p} = 0$, $P_{s} = \SI{2}{W.m^{-3}}$, $\tau_l/\tau_0^\phi = 10$ and $T_b/T_c = 0.1$.}
  \label{fig:5}
\end{figure}
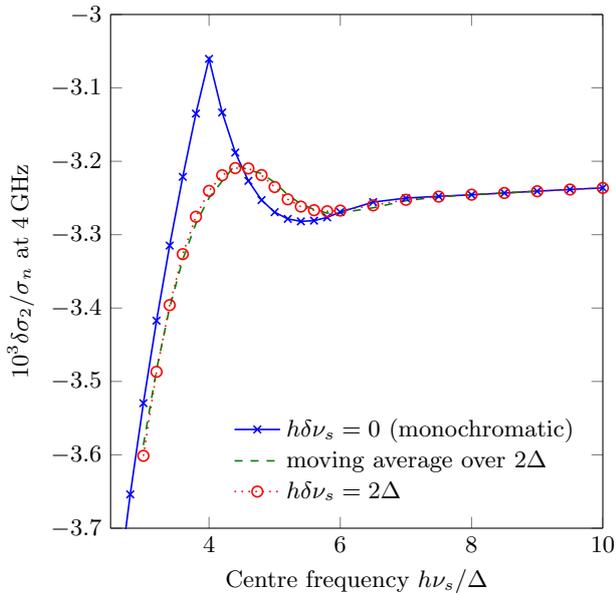

It is the changing surface impedance of the superconducting resonator of KIDs which is measured when they are operated as detectors.
Nonequilibrium structure in $f(E)$ near the gap energy can introduce small changes to the conductivity which cannot be accounted for by simply considering changes in overall quasiparticle number~\cite{Goldie2013,DeVisser2013a}, so in principle it would be possible that a broadband source causes changes in conductivity compared to a monochromatic source even though total quasiparticle numbers, and $\eta$, were equal.
Here however we have shown the low energy structure of $f(E)$ is essentially identical when considering broadband versus monochromatic sources, and therefore averaging simpler calculations assuming monochromatic sources over the signal bandwidth is sufficient for calculating distribution-averaged quantities after broadband absorption.
\Cref{fig:5} plots the change in the imaginary part of the complex conductivity $\delta \sigma_2$ when the superconductor is illuminated. This was calculated using the nonequilibrium $f(E)$ in the Mattis-Bardeen equations~\cite{Mattis1958}, and is given as a fraction of the normal state conductivity $\sigma_n$. Comparing the full calculation including a broadband source (red $\bigcirc$) to the moving average over the bandwidth $2\Delta$ (dashed green) of the monochromatic calculation (blue $\times$) shows excellent agreement.

\subsection{Readout (sub gap frequency) photons}

\begin{figure}[htb]
  \centering
  
  \ifhavepgf
  \tikzset{external/remake next}
  \tikzexternaldisable
  \newsavebox\mainplot
  \savebox\mainplot{\input{figures/letter_figure10P.tikz}}
  
  \newsavebox\insetplot
  \savebox\insetplot{%
  \footnotesize
  \setlength{\figureheight}{0.20\textwidth}%
  \setlength{\figurewidth}{0.18\textwidth}%
%
%
\begin{tikzpicture}

\begin{axis}[%
width=0.95092\figurewidth,
height=\figureheight,
at={(0\figurewidth,0\figureheight)},
scale only axis,
separate axis lines,
every outer x axis line/.append style={black},
every x tick label/.append style={font=\color{black}},
xmin=1,
xmax=1.5,
every outer y axis line/.append style={black},
every y tick label/.append style={font=\color{black}},
ymode=log,
ymin=1.9e-05,
ymax=0.0006,
yminorticks=true,
extra description/.code = {\node[coordinate,pin=above:{}] at (axis cs:1.09,0.0002) {};
\node[coordinate,pin=above:{}] at (axis cs:1.18,0.0002) {};
\draw [semithick,decoration={brace,raise=3pt},decorate] (axis cs:1.09,0.0003) -- node[above=5pt] {$h\delta \nu_p$} (axis cs:1.18,0.0003);}
]
\addplot [color=blue,solid,forget plot]
  table[row sep=crcr]{%
1	0.000189190887822972\\
1.005	0.000175505344081811\\
1.01	0.000174604891394544\\
1.015	0.000173563210763624\\
1.02	0.00017241346232414\\
1.025	0.000171170979730819\\
1.03	0.000169844445987398\\
1.035	0.000168439448878981\\
1.04	0.000166959931898177\\
1.045	0.000165408885528469\\
1.05	0.000163788715236645\\
1.055	0.000162101455943614\\
1.06	0.000160348908713928\\
1.065	0.000158532738278283\\
1.07	0.000156654555428046\\
1.075	0.000154716006197654\\
1.08	0.000152718903450852\\
1.085	0.000150665514794142\\
1.09	0.000166075937805792\\
1.095	0.000167024037758659\\
1.1	0.00016696733791885\\
1.105	0.000166562260831992\\
1.11	0.000165976996348596\\
1.115	0.000165274984300787\\
1.12	0.000164485547557675\\
1.125	0.000163624044170614\\
1.13	0.000162699214755013\\
1.135	0.000161716326029877\\
1.14	0.00016067867664278\\
1.145	0.00015958838160115\\
1.15	0.000158446807887716\\
1.155	0.000157254828407704\\
1.16	0.000156012975195605\\
1.165	0.000154721533673297\\
1.17	0.000153380603773123\\
1.175	0.000151990188158243\\
1.18	0.000151236713145389\\
1.185	0.000147269566892582\\
1.19	0.000144909776038545\\
1.195	0.000142927962805201\\
1.2	0.000141109961158688\\
1.205	0.000139379905362071\\
1.21	0.000137701091635867\\
1.215	0.000136052764174349\\
1.22	0.000134422062968047\\
1.225	0.000132800537826454\\
1.23	0.00013118240837063\\
1.235	0.000129563605040276\\
1.24	0.000127941200530307\\
1.245	0.000126313053093252\\
1.25	0.000124677572051203\\
1.255	0.000123033556890027\\
1.26	0.000121380082000264\\
1.265	0.00011971641630157\\
1.27	0.00011807304438989\\
1.275	0.000115958894126691\\
1.28	0.000113759887373174\\
1.285	0.000111519053226431\\
1.29	0.000109255127186782\\
1.295	0.000106979048424692\\
1.3	0.000104698185905938\\
1.305	0.00010241794640807\\
1.31	0.000100142533861429\\
1.315	9.78753592849796e-05\\
1.32	9.5619282614873e-05\\
1.325	9.33767643771827e-05\\
1.33	9.11499650447487e-05\\
1.335	8.89408121953881e-05\\
1.34	8.67510469381733e-05\\
1.345	8.45822565016506e-05\\
1.35	8.24358973092653e-05\\
1.355	8.03133116391858e-05\\
1.36	7.82169952663737e-05\\
1.365	7.61192722352254e-05\\
1.37	7.40684178162432e-05\\
1.375	7.20376140461218e-05\\
1.38	7.00214834498795e-05\\
1.385	6.80187779075032e-05\\
1.39	6.60298162029024e-05\\
1.395	6.40556806343148e-05\\
1.4	6.20978583738572e-05\\
1.405	6.01580500123523e-05\\
1.41	5.82380562685865e-05\\
1.415	5.63397065395107e-05\\
1.42	5.44648119719178e-05\\
1.425	5.26151338576999e-05\\
1.43	5.07923620745189e-05\\
1.435	4.89981003668231e-05\\
1.44	4.72338564409748e-05\\
1.445	4.55010355664844e-05\\
1.45	4.3800979436065e-05\\
1.455	4.21335756406736e-05\\
1.46	4.05061560846685e-05\\
1.465	3.89198896712275e-05\\
1.47	3.73750832193271e-05\\
1.475	3.58716779709418e-05\\
1.48	3.44094268627253e-05\\
1.485	3.29879786761222e-05\\
1.49	3.16069230402385e-05\\
1.495	3.0265815657871e-05\\
1.5	2.89641924849994e-05\\
1.505	2.7701577393596e-05\\
};
\addplot [color=black!50!green,dashed,forget plot]
  table[row sep=crcr]{%
1	0.000465370170688609\\
1.005	0.000215928065452462\\
1.01	0.000167692828395867\\
1.015	0.000143233979991406\\
1.02	0.000127400252166642\\
1.025	0.000115831147543919\\
1.03	0.000106749760005783\\
1.035	9.92778861290989e-05\\
1.04	9.29245011288951e-05\\
1.045	8.7390374128515e-05\\
1.05	8.24811650880921e-05\\
1.055	7.80643124335478e-05\\
1.06	7.40458054669723e-05\\
1.065	7.03568337222063e-05\\
1.07	6.69457031239462e-05\\
1.075	6.3772727919383e-05\\
1.08	6.08068876589297e-05\\
1.085	5.80235765918519e-05\\
1.09	5.54030570030119e-05\\
1.095	5.29293871299874e-05\\
1.1	5.05896909790002e-05\\
1.105	4.83737072063236e-05\\
1.11	4.62736286752277e-05\\
1.115	4.42843837876868e-05\\
1.12	4.24049275629032e-05\\
1.125	4.0642903747902e-05\\
1.13	3.90373594561919e-05\\
1.135	0.000455320974017715\\
1.14	0.000209553811882211\\
1.145	0.000161076366246538\\
1.15	0.000136492053879367\\
1.155	0.000120577431168151\\
1.16	0.000108952194699165\\
1.165	9.98315630984224e-05\\
1.17	9.23334687089486e-05\\
1.175	8.59647035663895e-05\\
1.18	8.04246657931405e-05\\
1.185	7.5518077382616e-05\\
1.19	7.11116865024088e-05\\
1.195	6.71109410451206e-05\\
1.2	6.34465813363623e-05\\
1.205	6.00665217061097e-05\\
1.21	5.6930719914033e-05\\
1.215	5.4007818343603e-05\\
1.22	5.12728811332545e-05\\
1.225	4.87058365055991e-05\\
1.23	4.62903928014545e-05\\
1.235	4.40132926940896e-05\\
1.24	4.18638376963537e-05\\
1.245	3.98336836222915e-05\\
1.25	3.79170302057379e-05\\
1.255	3.61116874561131e-05\\
1.26	3.44230411310169e-05\\
1.265	3.28835151176595e-05\\
1.27	0.000355032872726656\\
1.275	0.000163334878246111\\
1.28	0.000125038294094621\\
1.285	0.000105404098835152\\
1.29	9.2574263315295e-05\\
1.295	8.31305726588616e-05\\
1.3	7.56772856483134e-05\\
1.305	6.95229737444121e-05\\
1.31	6.42800080702124e-05\\
1.315	5.97114398191761e-05\\
1.32	5.56628312929983e-05\\
1.325	5.20284039255039e-05\\
1.33	4.87327809362796e-05\\
1.335	4.57204818992819e-05\\
1.34	4.2949554076981e-05\\
1.345	4.03875426512905e-05\\
1.35	3.80088488925599e-05\\
1.355	3.57929472191317e-05\\
1.36	3.37231547425975e-05\\
1.365	3.17857710684293e-05\\
1.37	2.99694804542223e-05\\
1.375	2.82649596586702e-05\\
1.38	2.66646842152075e-05\\
1.385	2.51630097283202e-05\\
1.39	2.37568453566553e-05\\
1.395	2.24482626323029e-05\\
1.4	2.12574282629419e-05\\
1.405	0.000217326522012794\\
1.41	9.94670656627703e-05\\
1.415	7.55574767829166e-05\\
1.42	6.31498607419171e-05\\
1.425	5.49651220667762e-05\\
1.43	4.88995539369194e-05\\
1.435	4.40914566141657e-05\\
1.44	4.01123890226765e-05\\
1.445	3.67211302680938e-05\\
1.45	3.37694937897019e-05\\
1.455	3.1160339367188e-05\\
1.46	2.88266937040865e-05\\
1.465	2.67204737825419e-05\\
1.47	2.48059888928621e-05\\
1.475	2.30559955748562e-05\\
1.48	2.14491974881747e-05\\
1.485	1.99686035262001e-05\\
1.49	1.86004173363125e-05\\
};
\end{axis}
\end{tikzpicture}%
%
  }
  \tikzexternalenable
  \fi
  
  \tikzsetnextfilename{letter_figure10P}
  \begin{tikzpicture}
    \node[anchor=south west, inner sep=0] (myplot) at (0,0) {\usebox\mainplot};
    \begin{scope}[x={(myplot.south east)}, y={(myplot.north west)}]
      \node[anchor=north east, inner sep=0] at (0.95,0.97) {\usebox\insetplot};
    \end{scope}
  \end{tikzpicture}

  \caption{Steady state quasiparticle distribution $f(E)$ under illumination by monochromatic (dashed green) or broadband (solid blue) readout photons.
  Calculated for Al, with $P_{p} = \SI{2E3}{W.m^{-3}}$, $P_{s} = 0$, $\tau_l/\tau_0^\phi = 1$ and $T_b/T_c = 0.1$.}
  \label{fig:10P}
\end{figure}

For sub gap photons, \cref{fig:10P} compares $f(E)$ between a monochromatic (dashed green) and a broadband (solid blue) source for moderate absorbed power ($P_p = \SI{E3}{W.m^{-3}}$). Here the $f(E)$ calculated for a monochromatic source shows multiple peaks starting at the energy gap $\Delta$ and separated by $h\nu_p = \SI{24}{\micro eV} = 0.135\Delta$. In the broadband case, the inset shows only a small step in $f(E)$ just above the gap. This significant difference in structure is explainable by considering that the same total power is absorbed in both cases, but in the broadband case the excess quasiparticle population is distributed over the bandwidth of the signal. We have previously shown that at high absorbed sub gap powers, the quasiparticle generation efficiency $\eta$ of the readout power decreases with power due to multiple photon absorption~\cite{Guruswamy2015b}.
In a nonequilibrium quasiparticle distribution where only a single photon produced peak is present, the average energy of the excited quasiparticles is $E = \Delta + h\nu_p$, and when these quasiparticles relax back towards the gap energy by emitting phonons which escape from the thin film, an energy $h\nu_p$ is lost, resulting in the quasiparticle generation efficiency $\eta < 1$.
If multiple peaks in $f(E)$ are present, when these excited quasiparticles relax back towards the gap, more energy is lost as phonons and so a lower $\eta$ (and lower $N_{qp}$) results.
Here, due to the reduced power at each frequency in the broadband case compared to the monochromatic case, a smaller fraction of energy is lost and there are a greater number of low energy excess quasiparticles.

\begin{figure}[htb]
  \centering
  \tikzsetnextfilename{letter_figure2P}
%
%
\definecolor{mycolor1}{rgb}{0.00000,0.75000,0.75000}%
\begin{tikzpicture}

\begin{axis}[%
width=0.95092\figurewidth,
height=\figureheight,
at={(0\figurewidth,0\figureheight)},
scale only axis,
separate axis lines,
every outer x axis line/.append style={black},
every x tick label/.append style={font=\color{black}},
xmin=0.005,
xmax=0.045,
xlabel={Bandwidth $h\delta\nu_p / \Delta$},
every outer y axis line/.append style={black},
every y tick label/.append style={font=\color{black}},
ymin=0.21,
ymax=0.29,
ylabel={$\eta$},
legend style={at={(0.97,0.03)},anchor=south east,legend cell align=left,align=left,fill=none,draw=none},
scaled x ticks = false,
x tick label style={/pgf/number format/.cd, fixed, precision=2}
]
\addplot [color=blue,solid,mark=x,mark options={solid}]
  table[row sep=crcr]{%
0.01	0.258583740991903\\
0.01	0.258619752650362\\
0.02	0.26708306923873\\
0.03	0.270659010995427\\
0.04	0.27237409525159\\
};
\addlegendentry{$h\nu_p = 0.1 \Delta$};

\addplot [color=black!50!green,dashed,mark=o,mark options={solid}]
  table[row sep=crcr]{%
0.005	0.233791679731786\\
0.015	0.251050322932432\\
0.025	0.259090997253235\\
0.025	0.259128316981931\\
0.035	0.263901947648259\\
0.045	0.266878820243627\\
};
\addlegendentry{$h\nu_p = 0.1325 \Delta$};

\addplot [color=red,dotted,mark=+,mark options={solid}]
  table[row sep=crcr]{%
0.01	0.237625543633684\\
0.01	0.23762513604733\\
0.02	0.249303195535973\\
0.03	0.256096744132634\\
0.04	0.260589143629656\\
};
\addlegendentry{$h\nu_p = 0.15 \Delta$};

\addplot [color=mycolor1,dash pattern=on 1pt off 3pt on 3pt off 3pt,mark=square,mark options={solid}]
  table[row sep=crcr]{%
0.005	0.216683548503796\\
0.015	0.234902720458678\\
0.025	0.243929702432506\\
0.025	0.243961163596261\\
0.035	0.249901579427104\\
0.035	0.249898298142235\\
0.045	0.254212638718601\\
};
\addlegendentry{$h\nu_p = 0.1775 \Delta$};

\end{axis}
\end{tikzpicture}%
  \caption{Quasiparticle generation efficiency $\eta$ as a function of readout bandwidth $\Delta h\nu_p$, with varying centre frequency.
  Calculated for Al, with $P_{p} = \SI{2E3}{W.m^{-3}}$, $P_{s} = 0$, $\tau_l/\tau_0^\phi = 1$ and $T_b/T_c = 0.1$.}
  \label{fig:2P}
\end{figure}

The fact that the quasiparticle distribution close to the gap is different for a broadband source compared to a monochromatic source, means there is a difference in distribution averaged quantities, unlike in the above gap frequency case.
Starting at total absorbed powers sufficient to make multiple photon absorption significant in $f(E)$ with a monochromatic source, as we increase the bandwidth we therefore see an increase in $\eta$, as the power absorbed at each frequency decreases. This is shown in \cref{fig:2P}. The limits approached by $\eta$ as bandwidth is increased are equivalent to $\eta_0$, the low power limits of $\eta$ described in~\refcite{Guruswamy2015b}.

\begin{figure}[htb]
  \centering
  \tikzsetnextfilename{letter_figure40P}
%
%
\definecolor{mycolor1}{rgb}{0.00000,0.75000,0.75000}%
\definecolor{mycolor2}{rgb}{0.75000,0.00000,0.75000}%
\begin{tikzpicture}

\begin{axis}[%
width=0.95092\figurewidth,
height=\figureheight,
at={(0\figurewidth,0\figureheight)},
scale only axis,
separate axis lines,
every outer x axis line/.append style={black},
every x tick label/.append style={font=\color{black}},
xmin=0.08,
xmax=0.2,
xlabel={Centre frequency $h\nu_p / \Delta$},
every outer y axis line/.append style={black},
every y tick label/.append style={font=\color{black}},
ymin=0.4,
ymax=0.65,
ylabel={$\eta$},
legend style={legend cell align=left,align=left,fill=none,draw=none},
scaled x ticks = false,
x tick label style={/pgf/number format/.cd, fixed, precision=2}
]
\addplot [color=blue,solid,mark=x,mark options={solid}]
  table[row sep=crcr]{%
0.09	0.600406881724443\\
0.1	0.575792536933846\\
0.11	0.552841898670313\\
0.12	0.53139668033659\\
0.135	0.501785289315626\\
0.145	0.483598511622644\\
0.155	0.466547454208669\\
0.165	0.450547389584212\\
0.18	0.428342184780045\\
};
\addlegendentry{$h\delta\nu_p = 0 \Delta$};

\addplot [color=black!50!green,dashed,mark=o,mark options={solid}]
  table[row sep=crcr]{%
0.0875	0.617373108349805\\
0.1125	0.553612551036154\\
0.1225	0.531374107625065\\
0.1325	0.510719333834173\\
0.1425	0.491498569859551\\
0.1575	0.465073039124109\\
0.1675	0.448889936180966\\
0.1775	0.433744667987095\\
};
\addlegendentry{$h\delta\nu_p = 0.005 \Delta$};

\addplot [color=red,dotted,mark=+,mark options={solid}]
  table[row sep=crcr]{%
0.09	0.613863149279627\\
0.1	0.58675169640187\\
0.11	0.561776789417332\\
0.135	0.507230958893018\\
0.145	0.48810376289045\\
0.155	0.470295085090177\\
0.165	0.453678020445864\\
};
\addlegendentry{$h\delta\nu_p = 0.01 \Delta$};

\addplot [color=mycolor1,dash pattern=on 1pt off 3pt on 3pt off 3pt,mark=square,mark options={solid}]
  table[row sep=crcr]{%
0.0875	0.626866244600501\\
0.1125	0.559653969880555\\
0.1225	0.536431265622851\\
0.1325	0.514969502204012\\
0.1675	0.451297149908491\\
0.1775	0.435816009037529\\
};
\addlegendentry{$h\delta\nu_p = 0.035 \Delta$};

\addplot [color=mycolor2,solid,mark=diamond,mark options={solid}]
  table[row sep=crcr]{%
0.0875	0.62833202577019\\
0.1125	0.560696164135883\\
0.1225	0.537344462317092\\
0.1325	0.515772300398583\\
0.1425	0.495791670032736\\
0.1575	0.468472469985888\\
0.1675	0.45182332529099\\
0.1775	0.436286220995642\\
};
\addlegendentry{$h\delta\nu_p = 0.045 \Delta$};

\end{axis}
\end{tikzpicture}%
  \caption{Quasiparticle generation efficiency $\eta$ as a function of centre readout frequency $h\nu_p$, with varying bandwidth of readout.
  Calculated for Al, with $P_{p} = \SI{2E-3}{W.m^{-3}}$, $P_{s} = 0$, $\tau_l/\tau_0^\phi = 1$ and $T_b/T_c = 0.1$.}
  \label{fig:40P}
\end{figure}
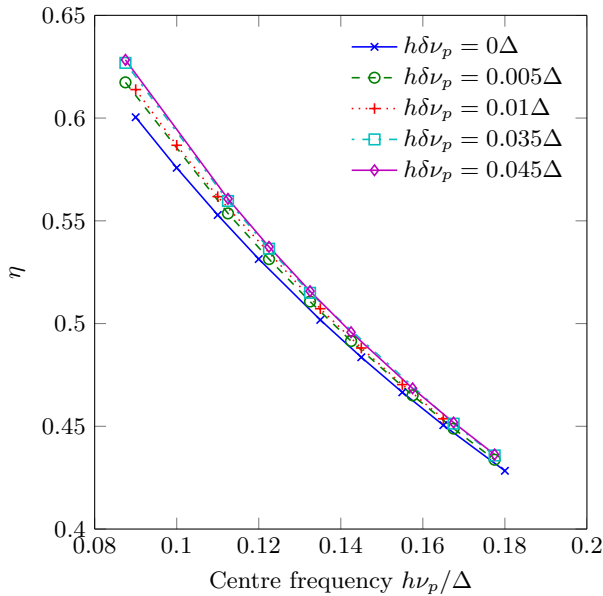

If the absorbed power is reduced significantly so that even for a monochromatic source no multiple photon absorption occurs, the bandwidth dependence of $\eta$ disappears. \Cref{fig:40P} shows an expected dependence of $\eta$ on centre frequency~\cite{Guruswamy2015b}, but no significant variation with bandwidth.

Typical Al KIDs designed for use in frequency multiplexed arrays have resonant frequencies on the order of $\SI{4}{GHz}$ and quality factors~\cite{Vardulakis2008} of the order of \num{E5}. Therefore, the bandwidth over which microwave absorption is significant is $\SI{E5}{kHz}$ or $\SI{E-10}{eV} \sim \num{E-6}\Delta$, much smaller than the bandwidths or even energy bin widths considered in this work. We would therefore not expect devices in these high-Q resonators to show evidence of the difference in quasiparticle generation between broadband and microwave sources.
However other applications of superconducting thin films, where broadband absorption is possible, may show significantly increased quasiparticle numbers when illuminated by a broadband source compared to a monochromatic one.

\section{Conclusions}

Existing models of KIDs assume a monochromatic signal to be detected.
However for many relevant KID applications, the signal (e.g. the CMB), sky noise, and thermal noise on the readout lines can all illuminate the detector with broadband radiation at both above gap and sub gap frequencies.
Therefore, we set out to check the implicit assumption that broadband absorbed power can be modelled as equivalent to a single tone with equal total absorbed power.
We calculate the steady state quasiparticle energy distributions in superconducting thin films under uniform illumination by broadband radiation sources.
When a superconducting thin film is illuminated by an above gap frequency source, the primary spectrum of excess quasiparticles generated by the absorbed photons is at high energies. Occupancy of these high energy states is low so absorption at multiple frequencies does not show any new features compared to absorption from a monochromatic source.
We show distribution averaged quantities such as quasiparticle generation efficiency $\eta$ and conductivity are identical to weighted averages (over the bandwidth of the source) of the corresponding quantities assuming a monochromatic source of the same total absorbed power.

In contrast, illumination by a sub gap source involves the creation of quasiparticles in states near the superconducting energy gap $\Delta$. For sufficient absorbed powers, quasiparticles can absorb multiple photons and be excited to high energies. A broadband source of the same total power as a monochromatic source has a smaller absorbed power at each frequency, so multiple photon absorption is reduced.
This results in significantly higher quasiparticle generation efficiency $\eta$ for a broadband source compared to a monochromatic source of the same moderate or high absorbed power.
For very low absorbed powers where multiple photon absorption is never significant, $\eta$ is the same for a broadband source and for a monochromatic source.
A thin film absorbing broadband microwave radiation will have a larger excess quasiparticle population than one absorbing the same total power at only one frequency.
Typically KIDs are designed with high quality factors and have sharp absorption characteristics, so absorption at a significant range of microwave frequencies should not occur, but this effect may be noticeable in other applications of superconducting thin films.

\bibliography{paper_broadband}   %
\end{document}